\documentclass{aip-cp}

\usepackage[numbers]{natbib}
\usepackage{rotating}
\usepackage{graphicx}

\newcommand{\sigmaSI}{\sigma_{\rm SI}}

\begin{document}

\title{SUSY Dark Matter: Beyond the Standard Paradigm}

\author[aff1]{Pearl Sandick\corref{cor1}}

\affil[aff1]{Department of Physics and Astronomy, University of Utah, Salt Lake City, UT 84112}
\corresp[cor1]{Corresponding author: sandick@physics.utah.edu}

\maketitle

\begin{abstract}
Within the framework of the Minimal Supersymmetric Standard Model (MSSM), we explore a decoupling of the parameters into separate sectors that determine consistency with collider data, the abundance of dark matter, and potential signatures at direct dark matter searches.  We consider weak-scale bino-like neutralino dark matter, and find that annihilations via light slepton exchange present a viable mechanism for obtaining the appropriate dark matter abundance assuming a thermal history.  Constraints and prospects for discovery of these models are discussed, including the possibility that direct dark matter searches may be sensitive to these models if light squarks exhibit left-right mixing.  Differences between the scenarios presented here and the typical expectations for the MSSM are discussed.
\end{abstract}

\section{INTRODUCTION}

Neutralino dark matter within the MSSM has long been among the most compelling candidates for particle dark matter.  In many regions of the MSSM parameter space, and in well-motivated, UV-complete subsets of the MSSM, the lightest neutralino is the lightest supersymmetric particle (LSP), and therefore assumed to be stable due to conservation of R-parity.  The abundance of the neutralino LSP, assuming that it's a thermal relic, is typically within a few orders of magnitude of the measured abundance of dark matter, $\Omega_{\chi}  h^2 = 0.1196 \pm{0.0031}$~\cite{Ade:2013zuv}, and in many cases obtains an abundance in perfect agreement with the measured value.  This promising scenario has led to much optimism regarding the discovery prospects for supersymmetry and supersymmetric dark matter.  However, no new supersymmetric particles have yet been discovered at the Large Hadron Collider (LHC), as no dark matter particles have yet been discovered via direct or indirect dark matter searches.  Null searches for supersymmetry and the discovery of the Higgs boson with a mass of $\sim 125$ GeV \cite{Higgs} seem to be pointing to the possibility that supersymmetric particles, if the Universe is indeed supersymmetric, are, in general, at the TeV scale or beyond.

However, the current data clearly leave open the possibility of electroweak-scale sleptons~\cite{PDG, Aad:2014vma}.  If thta is the case, as will be shown here, it is possible to satisfy all collider constraints and obtain a Higgs boson with a mass of $\sim 125$ GeV in models where the neutralino LSP is bino-like and a true weakly interacting massive particle (WIMP) dark matter candidate~\cite{FKKSY}.  Here we explore this scenario, the possibility that consistency with collider data and consistency with the dark matter abundance can be obtained by considering two separate sectors of the MSSM parameter space, as well as the related prospects for direct dark matter detection in these types of models~\cite{KKSS}.

\section{MODEL AND CONSTRAINTS}

While there are a plethora of observable quantities predicted within a supersymmetric model that, if different from the Standard Model (SM) expectation, would be evidence of supersymmetry or particle dark matter, there are currently no convincing observations of any deviations from the SM.  If we are driven by the desire for electroweak-scale WIMP dark matter in an MSSM scenario consistent with collider constraints, we recognize that a very simple scenario survives:  singlet fermion dark matter coupled to SM fermions via charged scalars.  Though this scenario, often termed the ``bulk'' region of parameter space, has long been excluded in supersymmetric extensions of the SM such as the constrained MSSM (CMSSM), in fact there is no conflict with collider data to date in the more general MSSM.  Here we re-open this region of parameter space for consideration, allowing a weak-scale bino dark matter candidate and sleptons with masses $\gtrsim100$ GeV~\cite{FKKSY} as allowed by current constraints~\cite{PDG, Aad:2014vma}.  

There are, however, well-known obstacles to obtaining an appropriate dark matter abundance with a bino-like LSP annihilating via light slepton exchange: namely, the fact that the $s$-wave part of the annihilation cross section is typically chirality-suppressed by a factor $\sim m_f^2 / m_\chi^2$, while the $p$-wave part of the cross section is velocity-suppressed by a factor $v^2 \approx 0.1$ at freeze-out.  Together, these suppressions imply a dark matter annihilation cross section much smaller than that required for the abundance to be consistent with measurements.  Allowing for significant left-right mixing in the slepton sector, however, eliminates the chirality suppression, restoring the viability of bulk-type models for explaining the observed abundance of dark matter.  In fact, one can completely specify the physics relevant to the relic density of dark matter by specifying the bino mass and slepton masses and mixings.  We note that we consider only annihilation via slepton exchange to SM leptons.  In principle, one could also consider neutralino-slepton coannihilation, which requires near degeneracy of the neutralino LSP with one of the sleptons, a much more ``tuned'' situation than that explored here.
In order to predict a light $CP$-even Higgs boson consistent with observations, there must also be a heavy sector of parameter space, which includes, at a minimum, the Higgs mixing parameter, $\mu$, some heavy squark masses, and the top trilinear couplings.  In practice the wino and gluino may also be decoupled, as well as some of the sleptons.  

Introducing left-right mixing in the slepton sector is not without consequence.  Contributions to the electric and magnetic dipole moments of the SM leptons arise as one-loop vertex corrections, with the bino and sleptons running in the loop.  Since dipole moment operators flip the lepton helicity, these contributions can be large if left-right slepton mixing is large.  Within the SM, the leading order contributions to the electric dipole moments of the charged leptons occur only at more than three loops~\cite{Booth:1993af} and are many orders of magnitude below the current constraints.  As the dipole moment contributions are dependent not only on the slepton mixing angle, but also on the $CP$-violating phase (the electric dipole moment vanishes in the absence of $CP$-violation), we allow a $CP$-violating phase for the sleptons, as well.  The relic density sector now also completely specifies the physics relevant for corrections to the lepton dipole moments, and includes the bino mass, $m_\chi$, and the slepton masses, $m_{\tilde{\ell}_{1,2}}$, mixing angles, $\alpha$, and $CP$-violating phases, $\varphi$.  For simplicity, we consider each generation of sleptons separately, but we note that while the contributions to the dipole moments are independent, the contributions to the annihilation cross section are additive.

\section{DARK MATTER ABUNDANCE AND LEPTON DIPOLE MOMENTS}

Because bino annihilation exhibits no $s$-channel resonances,
one can expand  $\langle \sigma v \rangle$ in powers of $T/m_{{\chi}}$~\cite{Srednicki:1988ce} as
\begin{equation}
\left \langle \sigma v \right \rangle \sim  c_0 + c_1 \left ( \frac{T}{m_{{\chi}}} \right ) ,
\end{equation}
where the second term on the right hand side is suppressed by the velocity squared, while $c_0$ is the velocity-independent $s$-wave contribution,
\begin{equation}
c_0 = {m_{\chi}^2 \over 2 \pi } g^4  Y_L^2 Y_R^2 \cos^2 \alpha \sin ^2 \alpha \left( {1 \over m_{\tilde \ell_1}^2
+ m_{\chi}^2} -{1 \over m_{\tilde \ell_2}^2  + m_{\chi}^2} \right) ^2,
\end{equation}
Here $g$ is the hypercharge coupling and $Y_{L,R}$ are the slepton hypercharges.  For simplicity, we have taken the limit of massless fermions, though the exact expressions for the annihilation cross section and the corrections to the dipole moments are used in the following analysis.  We see that the $s$-wave contribution to the annihilation cross section is maximized for maximal left-right mixing, and vanishes in the absence of mixing or when the sleptons are degenerate.

Again in the massless limit, the contributions to the anomalous magnetic moment, $a = \frac{g-2}{2}$, and the electric dipole moment,
$d/|e|$, of the associated lepton due to new physics are~\cite{Cheung:2009fc}
\begin{equation}
\Delta a =
 {m_{\ell} m_{\chi} \over 4 \pi^2 m_{\tilde \ell_1}^2} g^2 Y_L Y_R \cos \varphi
 \cos \alpha \sin \alpha   \left[ {1 \over 2(1- r_{1})^2} \left( 1+  r_{1} + {2 r_{1} \ln  r_{1} \over
 1- r_{1}}\right)   \right]
 -(\tilde \ell_1 \to \tilde \ell_2)
 \label{eq:mdm}
\end{equation}
and
\begin{equation}
 {d \over |e|} = {m_{\chi} \over 8 \pi^2 m_{\tilde \ell_1}^2 } g^2 Y_L Y_R \sin \varphi
 \cos \alpha \sin \alpha  \left[ {1 \over 2(1- r_{1})^2} \left( 1+  r_{1} + {2 r_{1} \ln  r_{1} \over
 1- r_{1}}\right)   \right]
-(\tilde \ell_1 \to \tilde \ell_2),
\label{eq:edm}
\end{equation}
where $ r_{i} \equiv m_{\chi}^2/m_{\tilde \ell_i}^2$.  The corrections to the dipole moments also vanish in the absence of left-right mixing or when the sleptons are degenerate.  Comparing Equation~\ref{eq:mdm} and Equation~\ref{eq:edm}, we see that they have identical dependence on $\alpha$, but the dependence on $\varphi$ is out of phase by $\pi/2$.

\begin{figure}[h!]
\begin{tabular} {c c}
(a)\, $\Omega_\chi h^2$ & (b)\, $(\sigma v)_{\rm ann.}$ today $/ (10^{-26} \,{\rm cm}^3/{\rm s})$ \\
\includegraphics[width=.45\textwidth]{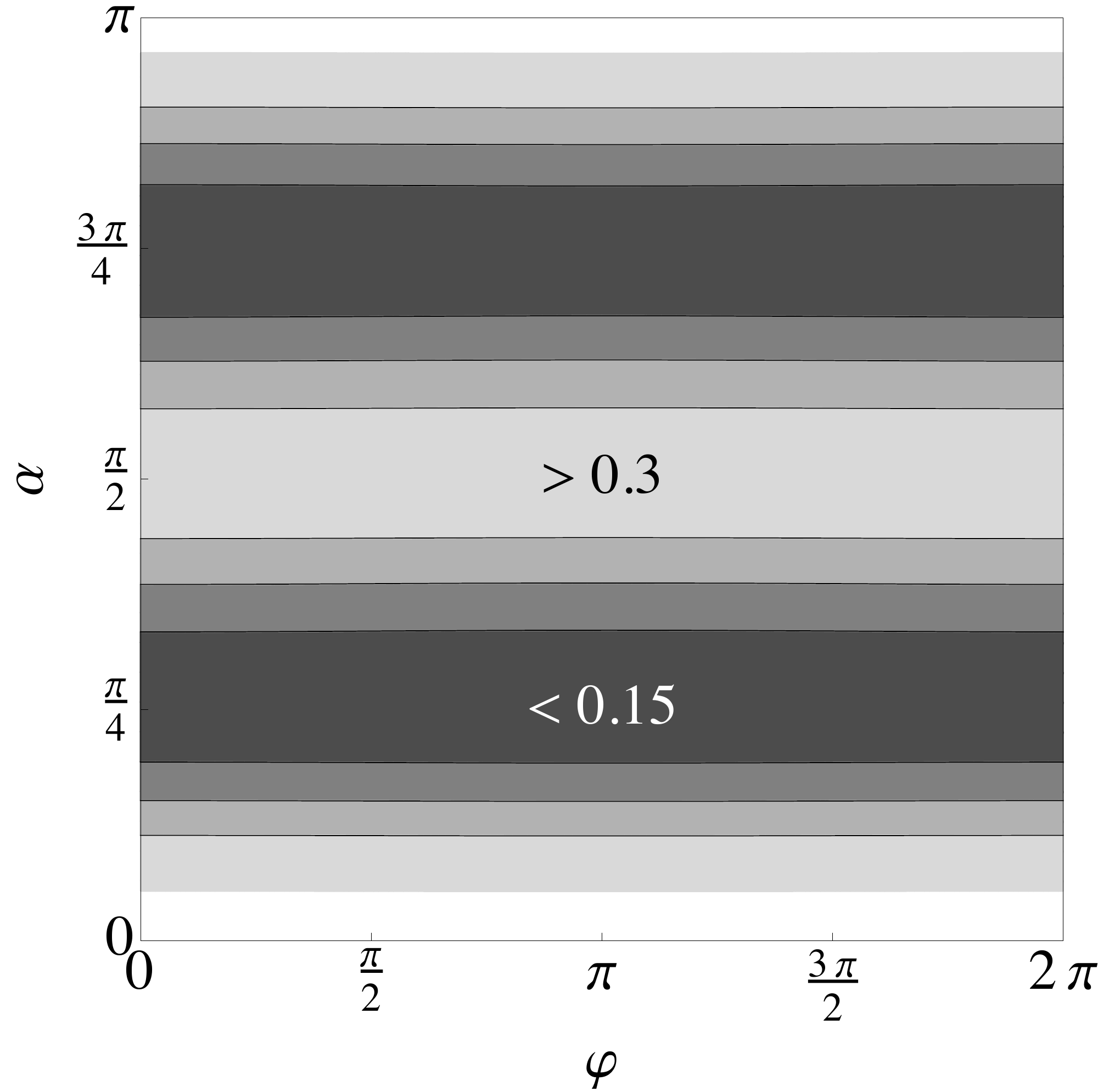} &
\includegraphics[width=.45\textwidth]{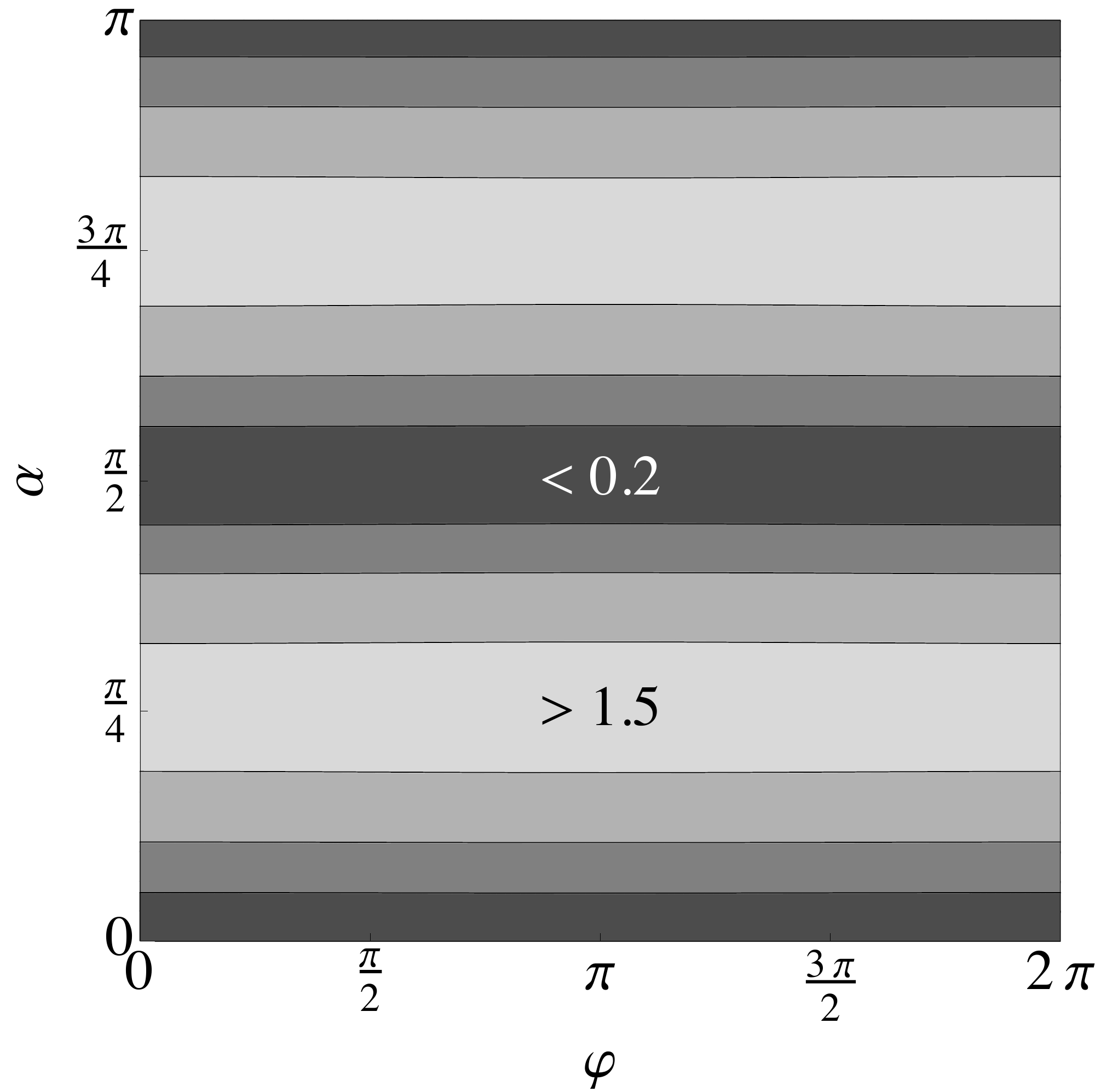}  \\[10pt]
(c)\, $\Delta a$ & (d)\, $2m_\mu d_\mu \,/ \left| e \right| $ \\
\includegraphics[width=.45\textwidth]{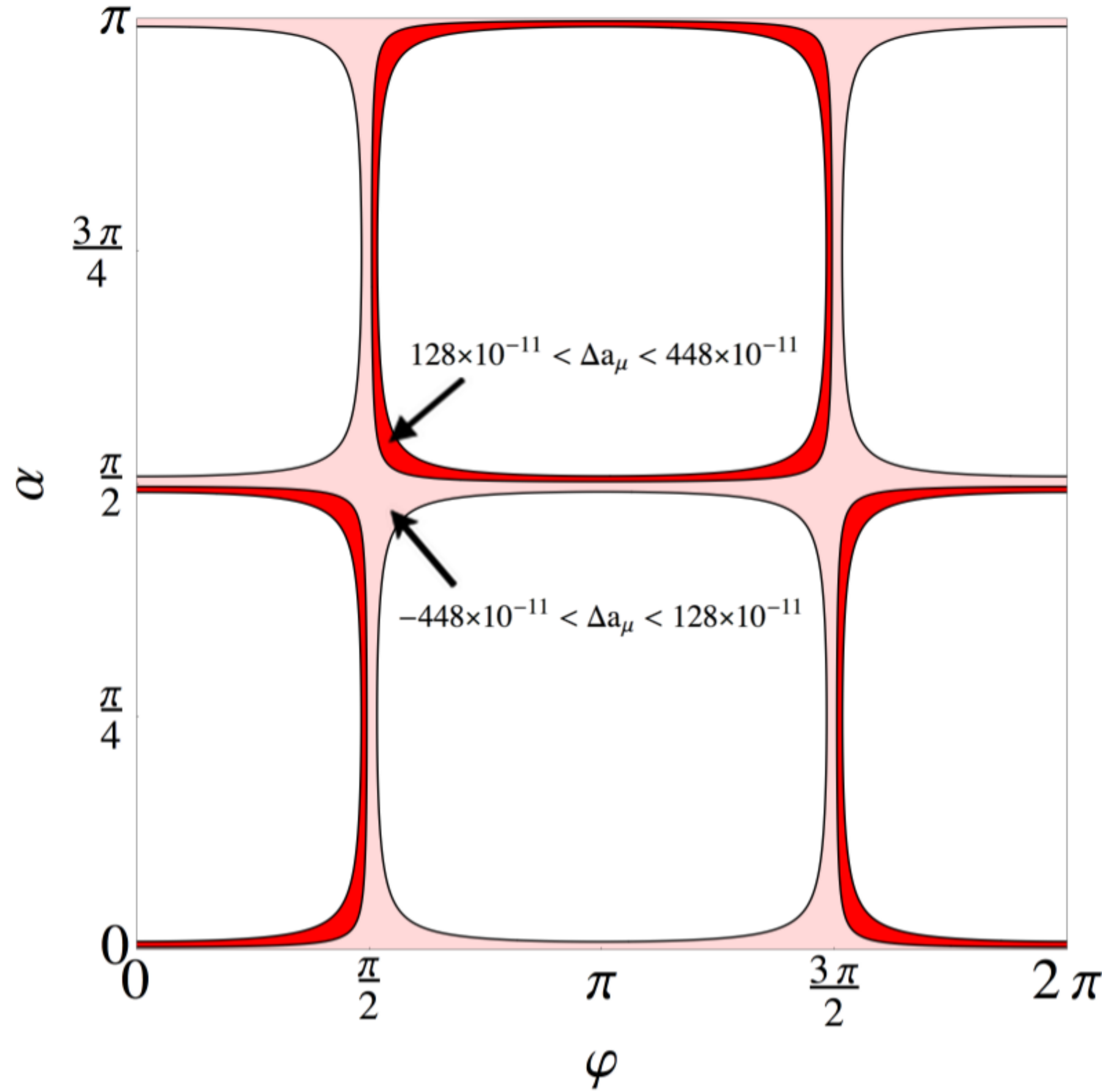} &
\includegraphics[width=.45\textwidth]{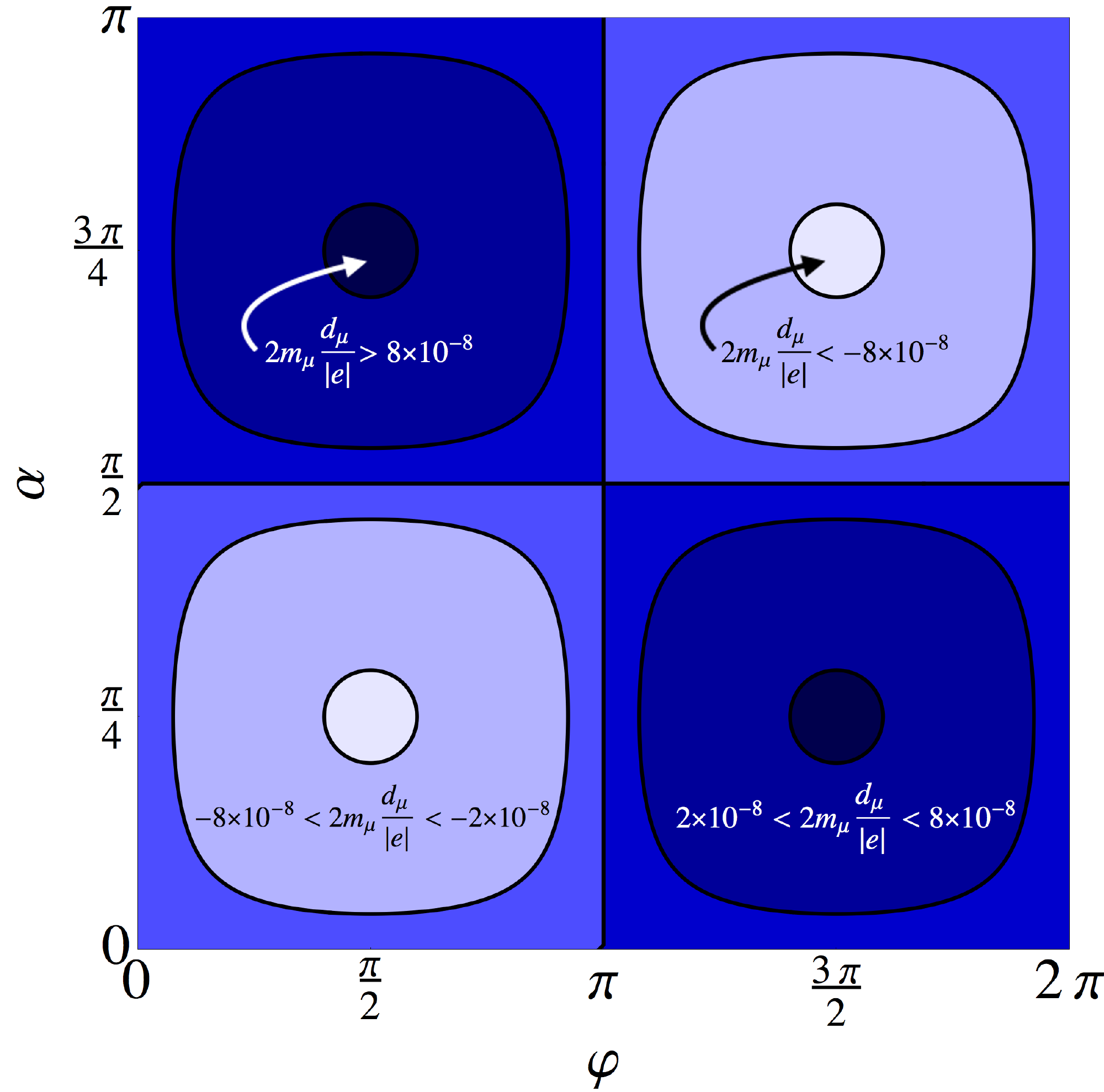} 
\end{tabular}
 \caption{
 \label{fig:basicbehavior}
The dependence of various observables on the smuon L-R mixing angle, $\alpha$, and the CP-violating phase, $\varphi$, for
$m_{{\chi}} = 100$ GeV, $m_{\tilde{\mu}_1}=120$ GeV, and $m_{\tilde{\mu}_2} = 300$ GeV.
In the upper two panels, we present the neutralino relic abundance (left) and the neutralino annihilation cross section today (right).  In the lower panels, we present the
contribution to the anomalous magnetic (left) and electric (right) dipole moments of the muon.  In the darker red region in the lower left panel
this model fully accounts for the measured muon anomalous magnetic moment to $2\sigma$, while the lighter red shaded region provides a contribution
that is comparable to the measured value in magnitude.  In the lower right panel,
the electric dipole moment is unconstrained everywhere in the plane.}
\end{figure}

In Figure~\ref{fig:basicbehavior} we present the dependence of the relevant constraints on the mixing angle, $\alpha$, and the $CP$-violating phase, $\varphi$, for a benchmark model with $m_\chi = 100$ GeV and light smuons with masses $m_{\tilde{\mu}_1} = 120$ GeV and $m_{\tilde{\mu}_2} = 300$ GeV.  The upper panels of Fig.~\ref{fig:basicbehavior} show (a) the neutralino relic abundance and (b) the annihilation cross section today, with darker shading in each panel indicating a smaller value.  As expected, larger mixing leads to a larger annihilation cross section and therefore a smaller relic abundance of neutralinos, becoming consistent with the measured abundance of dark matter near maximal mixing ($\alpha \approx \pi/4$).  We note that the dependence on $\varphi$ appears only in terms proportional to the fermion mass, and is therefore only marginally relevant even for annihilation to $\tau^+\tau^-$, when it can be as large as a $5\%$ effect.
The lower panels of Figure~1 show the constraints on (c) the anomalous magnetic moment and (d) the electric dipole moment of the muon.  In panel (d), darker shading indicates a larger correction, though all points in the plane lie well below the sensitivity of current measurements of the electric dipole moment of the muon~\cite{Bennett:2008dy, Booth:1993af}.  In panel (c), the darker shading indicates the region in which the discrepancy in the measurement of the muon anomalous magnetic moment~\cite{muon} is resolved, while the lighter shaded region yields a contribution that is comparable in magnitude to the current discrepancy (that is, the problem is not significantly worsened in this region, while it is significantly worsened in the unshaded regions of the plane).  As expected, the contributions to the lepton dipole moments vanish in the absence of left-right mixing.  For this benchmark point, the abundance of dark matter can be obtained and the anomalous magnetic moment of the muon can be explained for nearly maximal left-right smuon mixing and $CP$-violating phase of $\varphi \approx n \pi/2$ for $n$ odd.

We note that the annihilation cross sections today displayed in the upper right panel of Figure~\ref{fig:basicbehavior} are quite close to the typical ``thermal cross section'' of $(\sigma v)_{\rm th} = 3 \times 10^{-26}$ cm$^3$s$^{-1}$, which are beginning to be probed by indirect dark matter searches.  Annihilation of $100$ GeV WIMPs to $\mu^+\mu^-$ with $(\sigma v) \gtrsim 2\times 10^{-26}$ cm$^3$s$^{-1}$ is currently excluded by AMS-02~\cite{Bergstrom:2013jra}, while Fermi's observations of dwarf galaxies put a strong constraint on annihilation to $\tau^+\tau^-$, which reaches near $(\sigma v)_{\rm th}$~\cite{Ackermann:2015zua}.  Unfortunately, indirect dark matter searches suffer from substantial astrophysical uncertainties, resulting in possible fluctuations of these constraints by a factor of a few in either direction.  Nonetheless, indirect dark matter searches may indeed provide the first hints of these models, or may exclude them within the next few years.

\begin{figure}[h]
\begin{tabular} {c c c}
(a) light selectrons & (b) light smuons & (c) light staus \\ 
\includegraphics[width=5.3cm]{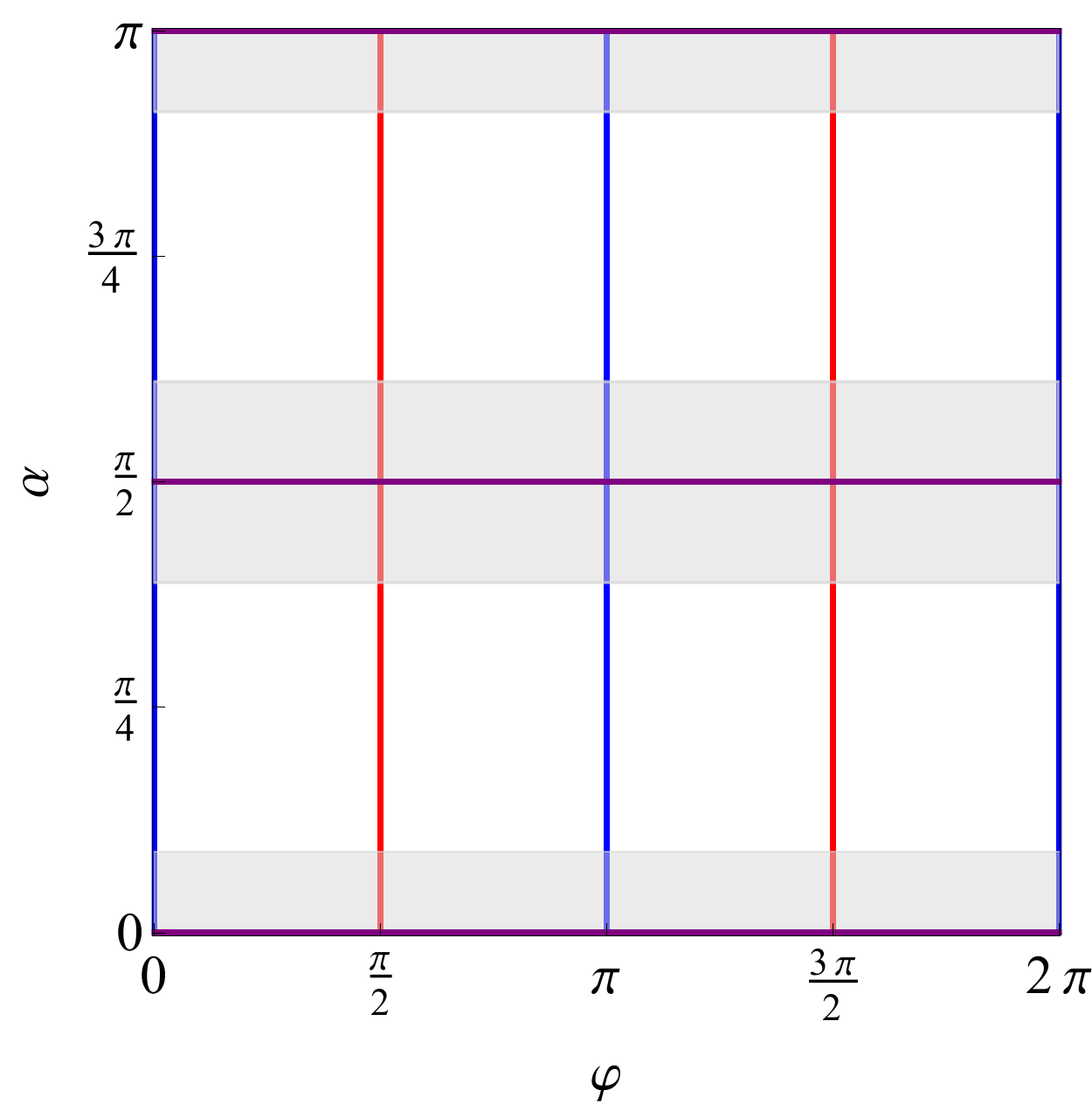} &
\includegraphics[width=5.3cm]{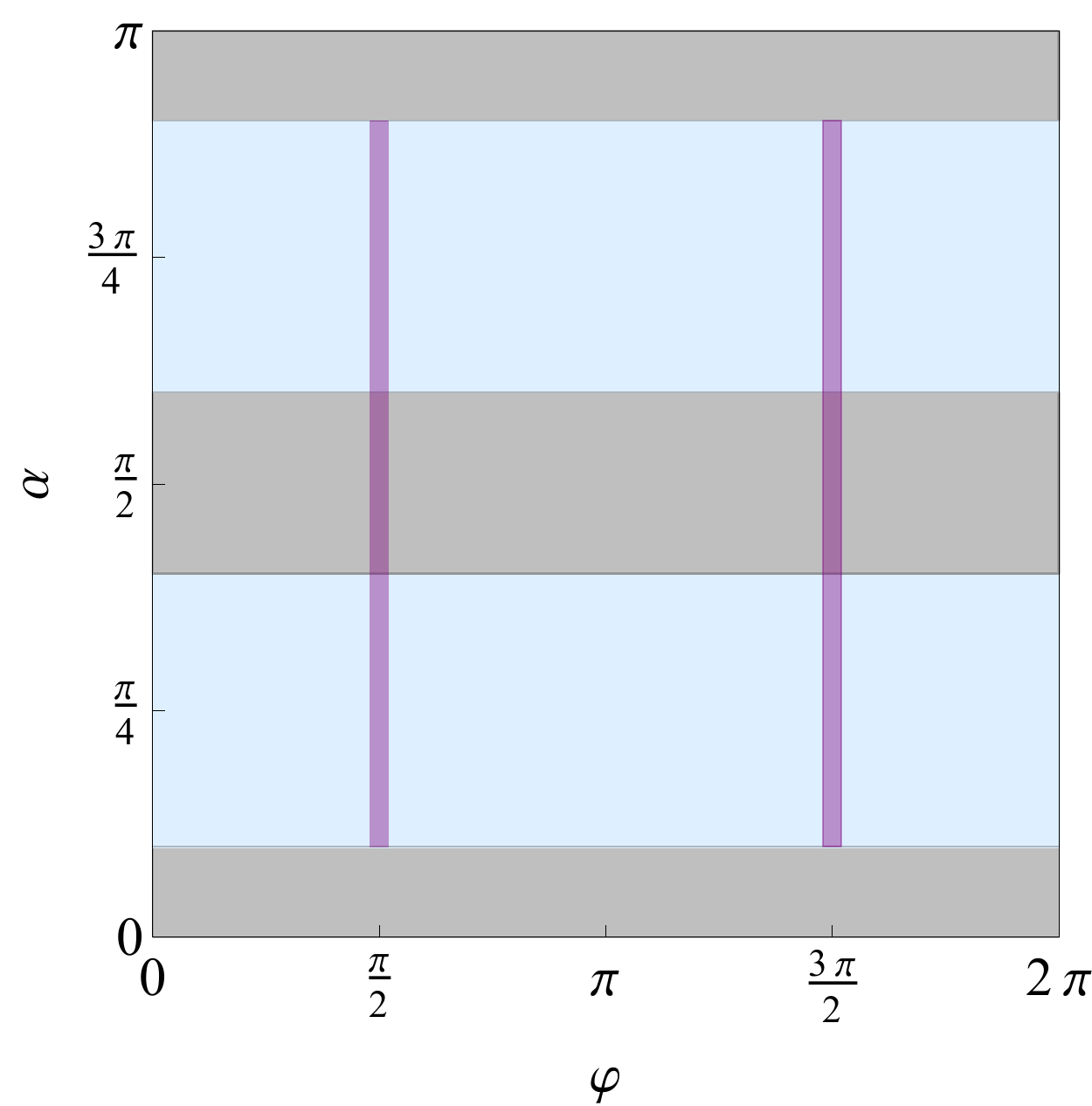} &
\includegraphics[width=5.3cm]{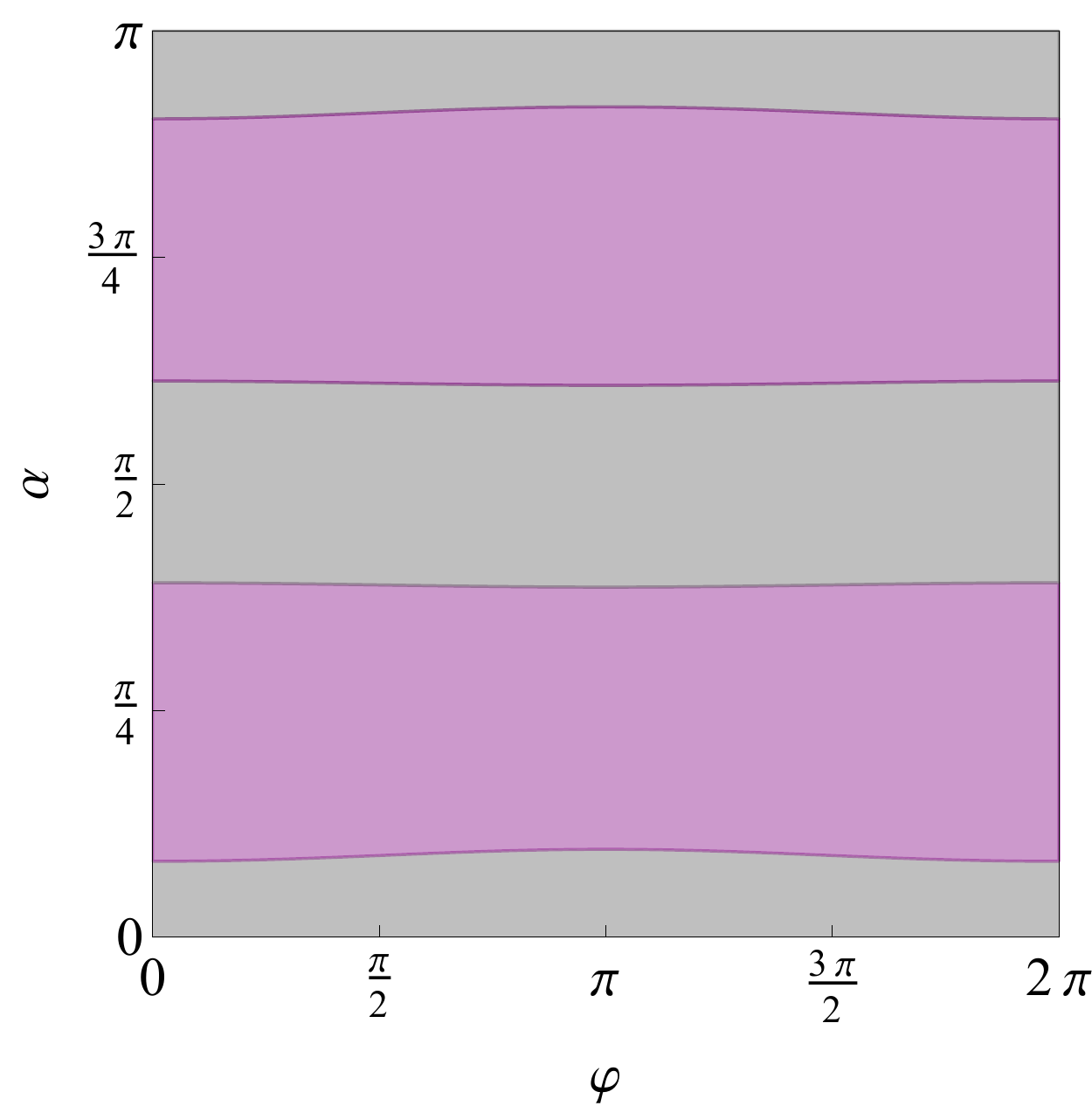}
\end{tabular}
 \caption{
 \label{fig:anglePlots}
A summary of the constraints for annihilation via (a) selectron exchange, (b) smuon exchange, and (c) stau exchange for $m_\chi=100$ GeV.
Greyed regions are excluded because the relic density of neutralinos would exceed the measured dark matter abundance
by more than $2\sigma$, while the blue/red shaded regions indicate angles where at least one $(m_{\tilde \ell_1}, m_{\tilde \ell_2})$ mass
combination produces an electric/magnetic dipole moment within the
current bounds. Purple regions indicate that both dipole moment constraints are simultaneously satisfied.
}
\end{figure}

In Figure~\ref{fig:anglePlots}, we present a summary of the constraints on these models for annihilation via (a) selectron exchange, (b) smuon exchange, and (c) stau exchange, in each case with sleptons of the other two generations decoupled, where $m_\chi = 100$ GeV and we have marginalized over $m_{\tilde{\ell}_1}$ and $m_{\tilde{\ell}_2}$.  In each panel, the grey regions are excluded because the relic density of neutralinos would exceed the measured dark matter abundance
by more than $2\sigma$, while the red and blue shaded regions are allowed by constraints on the magnetic and electric dipole moments, respectively, for at least one combination of $m_{\tilde \ell_1}$ and $m_{\tilde \ell_2}$.  In all cases, thermal neutralino dark matter with $m_\chi = 100$ GeV can make up the dark matter in the Universe if left-right slepton mixing is significant.  It is clear that the dipole moment constraints on light selectrons eliminate selectron exchange as a viable mechanism for obtaining the appropriate dark matter abundance.  Even with another mechanism by which to obtain the dark matter abundance, $\alpha$ and $\varphi$ must be highly tuned if selectrons are light.  The situation for light smuons is quite different: the muon electric dipole moment does not constrain $\alpha$ or $\varphi$, but measurements of the anomalous magnetic dipole moment of the muon require $\varphi \approx n\pi/2$, with $n$ odd.  Neither the magnetic nor the electric dipole moments of the tau lepton constrain the parameter space for the scenario with light staus, leaving $\varphi$ free.  

This scenario exhibits a few important departures from the standard expectations for neutralino dark matter within the MSSM.  Most importantly, a thermal abundance of neutralino dark matter can be obtained via exchange of light scalars, i.e.~the ``bulk'' mechanism.  We have obtained this result in a scenario with a simplified relic density sector, allowing the satisfaction of all other constraints by invoking a heavy sector of supersymmetric parameters.  This model can therefore be considered a {\it minimal} version of the MSSM with leptophilic dark matter.  

Regarding detectability, this region of the MSSM parameter space gives rise to many distinctive signatures.  Specifically, indirect dark matter searches may soon probe these models.  Lepton dipole moment constraints imply that only for annihilation to $\mu^+\mu^-$ and $\tau^+\tau^-$ can the annihilation cross section today be large enough for a potential signal in an indirect dark matter search, and, furthermore, for annihilation to $\mu^+\mu^-$ that would imply large $CP$-violation in the smuon sector.  We find that this mechanism only works for mediating sleptons lighter than $\sim150$ GeV, making this a well-defined region with quite small slepton masses for collider searches to target.  However, with typical mass splittings of $m_{\tilde{\ell}_1} - m_\chi \sim 10$ to 50 GeV, this may be a particularly challenging scenario to probe.  If annihilation proceeds primarily to $\tau^+\tau^-$, the branching fraction of the light CP-even Higgs to photons may be affected~\cite{Pierce:2013rda}.  Interestingly, if the mass of the lightest slepton can be constrained to be larger than $\sim150$ GeV, then it would {\it not} be possible to explain the observed dark matter abundance without invoking coannihilations or resonant annihilations in the early Universe and/or a non-trivial wino/Higgsino fraction for the LSP.

\section{DIRECT DARK MATTER SEARCHES}

If the neutralino LSP is a pure bino, scattering with nuclei can happen solely via squark exchange.  In the absence of left-right mixing in the squark sector, this scattering is either spin-dependent or velocity-suppressed. In either case, it is unlikely to be observed in direct dark matter searches.  However, if there is substantial left-right squark mixing, there is a velocity-independent contribution to the spin-independent neutralino-nucleon elastic scattering cross section.  Thus far, we have considered squarks to be decoupled so as to satisfy collider constraints, however these constraints still leave open the possibility that some squarks are quite light.  For one non-degenerate light flavor squark and $m_\chi = 100$ GeV, $m_{\tilde{q}_1} \gtrsim 500$ GeV~\cite{Aad:2014wea, Mahbubani:2012qq}.  In this case, the relic abundance of dark matter, if obtained via the mechanism described above, would be unaffected, but significant scattering with nuclei may be expected.  Furthermore, if the neutralino LSP is significantly heavier than 100 GeV such that slepton exchange is not adequate to achieve the measured abundance of dark matter, it is still possible that the abundance is achieved through neutralino-squark coannihilations.  In this section we briefly investigate neutralino-nucleon scattering, which we may describe with a third decoupled sector of the MSSM parameter space, specified by the bino mass, $m_\chi$, the light squark masses, $m_{\tilde{q}_{1,2}}$, and the squark left-right mixing angle, $\phi_q$~\cite{KKSS}.  As above, we consider the cases of light up, down, and strange squarks separately.

For the momentum transfer relevant for WIMP-nucleon elastic scattering, the scattering operator can be represented as a WIMP-quark four-point contact interaction, leading to the well-known result for the elastic scattering cross section~\cite{SigmaSI}
\begin{equation}
\sigmaSI^N = {  \mu^2 \over 4 \pi  }
g^{\prime 4} Y_L^2
\left\{ \sum_{q} \sin (2\phi_{\tilde{q}}) Y_{Rq} \left[ {1 \over (m_{\tilde{q}_1}^2 - m_{\tilde{\chi}}^2) } - {1 \over (m_{\tilde{q}_2}^2 - m_{\tilde{\chi}}^2) } \right]
(B_q^N) \lambda_q \right\}^2,
\label{eq:SI_cross_section}
\end{equation}
where, as above, $g$ and $Y_{L,R}$ are the hypercharge coupling and hypercharges of the left- and right-handed quarks, $\mu$ is the reduced mass of the WIMP-nucleon system, $B_q^{N}$ are the integrated
nucleon form factors for quarks $q=u,d,s$ and nucleons $N=p,n$, and the factor $\lambda_q$ accounts
for the running of the scattering operator ${\cal O}$ from the weak scale ($\sim m_Z$) to the nucleon scale $\mu$.  Three independent parameters are necessary to specify the light quark content of the nucleon; the proton or neutron form factors,
\begin{equation}
B_u^{n} = B_d^{p}, B_d^{n} = B_u^{p}, \textrm{ and } B_s^{n} = B_s^{p},
\end{equation}
or, perhaps more conveniently, the parameters
\begin{eqnarray}
&\Sigma_{\pi N} \equiv \frac{\displaystyle m_u + m_d}{\displaystyle 2} \left( B_u^{N} + B_d^{N} \right), \\ \nonumber
&\sigma_0 \equiv \frac{\displaystyle m_u+m_d}{\displaystyle 2} \left( B_u^{N}+B_d^{N}-2B_s^{N} \right), \\
&z \equiv {\displaystyle B_u^p-B_s^p \over \displaystyle B_d^p-B_s^p}, \nonumber
\end{eqnarray}
where $\Sigma_{\pi N}$ can be determined from pion-nucleon scattering data, but with large uncertainties, $\sigma_0$ can be fit from the baryon masses in chiral perturbation theory or inferred from lattice QCD studies, but with large discrepancies between methods, and $z$ is calculated purely from baryon octet mass differences, with the result $z=1.49$ with negligible uncertainties.  To summarize, the values of the $B_q^N$ in Equation~\ref{eq:SI_cross_section} are highly uncertain.  Furthermore, we note that it is only in the case of light strange squarks, with up and down squarks decoupled, that dark matter will scatter with the same cross section on protons and neutrons ($B_s^p = B_s^n$).  In all other cases, scattering will be generically ``isospin violating,'' and limits from direct dark matter searches must be recalculated rather than directly applied.

\begin{figure}
\begin{tabular} {c c}
(a) Sensitivity to Strangeness Content of Nucleon & (b) Squark Mass Reach \\
\includegraphics[height=6cm]{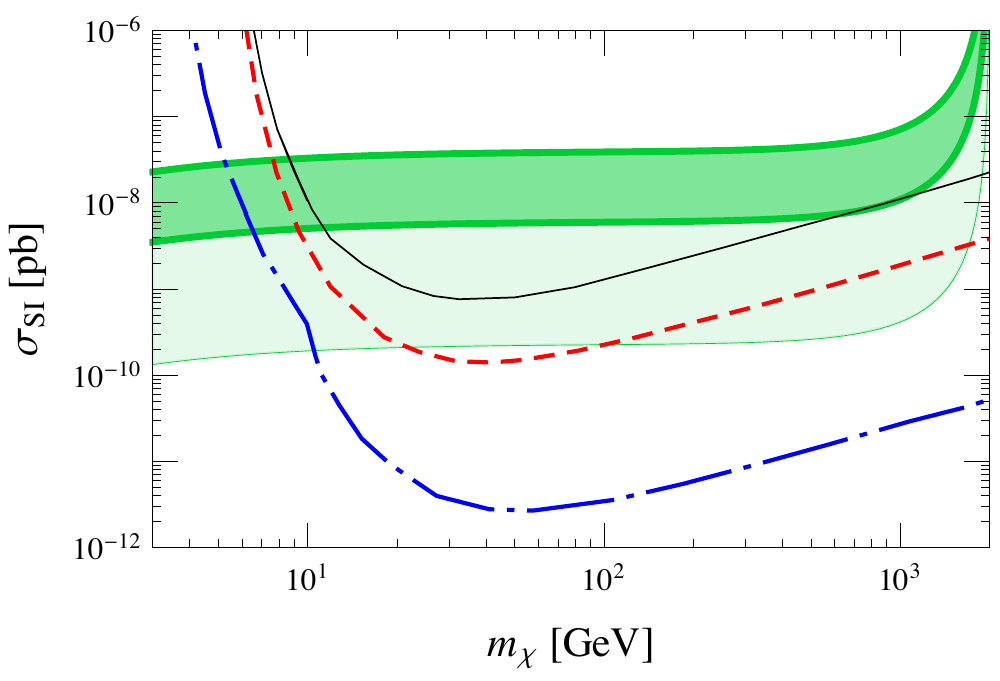} \hspace{3mm} &
\includegraphics[height=5.8cm]{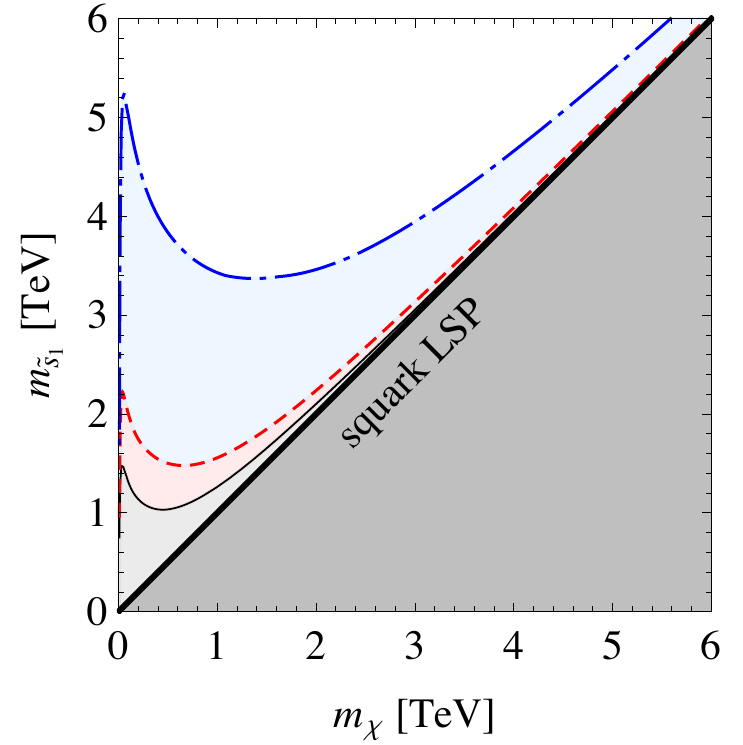}  \hspace{2mm} 
\end{tabular}
\caption{ Direct detection sensitivity for a benchmark case in which bino-like neutralinos scatter with nuclei via strange squark exchange.  Here we assume only one light squark, $m_{\tilde{s}_1}$, decoupling the heavier strange squark with $m_{\tilde{s}_2}=10$ TeV as well as the other squarks, and maximal left-right mixing, $\sin(2\phi_{\tilde{s}})=1$.  Panel (a) shows the $(m_{{\chi}}, \sigmaSI^N)$ plane for $m_{\tilde s_1} =2$ TeV.  The black line is the current upper limit from LUX~\cite{LUX}, while the dashed red line is the sensitivity curve for LUX with 300 days of data, and the
blue dot-dashed curve will be probed by LZ-7.
The green bands indicate the uncertainty in the scattering cross section as a function of the strangeness content
of the nucleon as described in the text.  
Panel (b) shows the $(m_{{\chi}}, m_{\tilde s_1})$ plane, assuming the minimal reference value $B_s^N =0.5$.  The grey region (between the solid line and squark LSP line) is ruled out by
current LUX data, while the red region (between the solid and dashed lines) could be probed by LUX with 300 days of data, and the
blue region (between the dashed and dot-dashed lines) will be probed by LZ-7.  
\label{fig:SIscattering}}
\end{figure}

In Figure~\ref{fig:SIscattering} we present the direct detection sensitivity for a benchmark case in which bino-like neutralinos scatter with nuclei via strange squark exchange.  Here we allow only one light squark, $m_{\tilde{s}_1}$, decoupling the heavier strange squark with $m_{\tilde{s}_2}=10$ TeV, and assume maximal mixing, $\sin(2\phi_{\tilde{s}})=1$ (note that results scale with $\sin(2\phi_{\tilde{s}})$ as in Equation~\ref{eq:SI_cross_section}).  In Fig.~\ref{fig:SIscattering}(a), we plot the spin-independent neutralino-nucleon elastic scattering cross section as a function of $m_\chi$ to illustrate the dependence of the predicted cross section on the form factors, $B_s^N$.  The green shaded region demonstrates this variation of the predicted scattering cross section:  The dark shaded band indicates the $2\sigma$ uncertainty in the measured value of $\Sigma_{\pi N}$~\cite{Alarcon:2011zs} assuming a minimal value of $\sigma_0=27$ MeV, while the light shaded band extends down to $B_s^N = 0.5$, at which point the strange quark content of the nucleon is equivalent to the heavy quark content, and, as such, one would not expect $B_s^N$ to be smaller.  We note that this ``minimal'' value of $B_s^N$ is in fact larger than the value favored by lattice calculations and the default value used by micrOMEGAS versions 3.0 through 3.5.5~\cite{Belanger:2013oya}.  The uncertainty in the scattering cross section due to the uncertain quark content of the nucleon extends over two orders of magnitude.  We also plot the current upper limit on the WIMP-nucleon scattering cross section from LUX~\cite{LUX} as a solid black contour, the projected sensitivity of LUX with 300 days of data as a red dashed contour, and the projected sensitivity of the LZ-7 experiment~\cite{LUXproj} as a blue dot-dashed contour.  One can see that conclusions regarding the detectability of a particular model are strongly dependent on the strange quark content of the nucleon.

In Figure~\ref{fig:SIscattering}(b) we plot the sensitivity of LUX and LZ in the $(m_\chi,m_{\tilde{s}_1})$ plane.  The lower right half of the plane is excluded as there would be a squark LSP.  The region below the thin black contour is currently excluded by LUX, while the red dashed and blue dot-dashed contours indicate projected sensitivities as in the left panel.  Here we have fixed $B_s^N=0.5$, the smallest value we would reasonably expect.  Even in this case, LZ will be sensitive to scattering via squark exchange so long as $m_{\tilde{s}_1} \gtrsim 3.2$ TeV.  In fact, for $m_\chi \approx 100$ GeV, LZ will be sensitive to scattering for squarks lighter than $\sim5$ TeV, and potentially to even heavier squarks if the light squark and neutralino LSP are nearly degenerate.  Even for the conservative value of $B_s^N=0.5$, direct dark matter searches will be sensitive to models in which the lightest colored superpartners are multi-TeV squarks with masses potentially well beyond the reach of the LHC operating at a center-of-mass energy of 14 TeV~\cite{CMSfuture}.  If heavy QCD-charged particles couple to dark matter, the first evidence may come from direct dark matter searches.

Again, this scenario contains many important departures from the standard MSSM expectation.  We have introduced an explicit scenario where squark exchange dominates spin-independent elastic scattering of neutralino dark matter with nuclei, whereas in the MSSM it is typical that Higgs-exchange dominates.  In the MSSM it is usually assumed that left-right mixing is negligible for light squarks (i.e.~minimal flavor violation), so that the scattering is basically isospin invariant.  Here, however, the scattering is generically isospin-violating, with the exception of the case where only strange squarks are light enough to participate significantly in the scattering.

\section{CONCLUSIONS}

While searches for supersymmetry and dark matter continue to return null results, we must look closely at models that evade constraints.  There are viable models within the MSSM with electroweak-scale neutralino dark matter, many of which exhibit distinctive signatures in colliders and direct and indirect dark matter searches.  Here, we have discussed two such scenarios, each with singlet fermion dark matter (a bino-like neutralino) coupled to SM fermions via charged scalars.  A bulk-like model with light sleptons and a $\sim100$ GeV bino-like neutralino LSP may explain the abundance of dark matter, while scattering with nuclei may very well proceed via squark exchange, each scenario enabled by left-right mixing among the relevant sleptons or squarks.  In the former case, indirect dark matter searches may be the key, while in the latter case, direct dark matter searches may be the first to provide evidence of physics beyond the SM (and will be significantly more powerful tools with an improved understanding of the quark content of the nucleon).


\section{ACKNOWLEDGMENTS}
The work of P.~Sandick is supported in part by NSF Grant No.~PHY-1417367.  P.S.~would also like to thank her collaborators on these projects and others, K.~Fukushima, C.~Kelso, J.~Kumar, P.~Stengel, and T.~Yamamoto.


\bibliographystyle{aipnum-cp}%

\end{document}